\documentclass[a4paper,UKenglish]{lipics-v2016}
\usepackage{numprint}
\usepackage{amsmath}
\usepackage{nccmath}
\usepackage{pdflscape}
\usepackage{lipsum}
\usepackage{wrapfig}
\usepackage{microtype}
\usepackage{graphicx}

\newcommand{\sodass}{\,:\,}
\newcommand{\setGilt}[2]{\left\{ #1\sodass #2\right\}}
\newcommand{\Is}       {:=}
\newcommand{\set}[1]{\left\{ #1\right\}}

\newcommand{\ie}{i.e.\ }
\newcommand{\etal}{et~al.~}
\newcommand{\eg}{e.g.\ }

\usepackage{tikz,pgfplots}
\usepgfplotslibrary{colorbrewer}
\usetikzlibrary{patterns}
\npdecimalsign{.}

\usepackage[algo2e,linesnumbered,ruled,lined]{algorithm2e}

\newcommand{\niceremark}[3]{\textcolor{red}{\textsc{#1 #2: }}\textcolor{blue}{\textsf{#3}}}

\newcommand{\alex}[2][says]{\niceremark{Alex}{#1}{#2}}
\newcommand{\csch}[2][says]{\niceremark{Christian}{#1}{#2}}

\renewcommand{\alex}[2][says]{}
\renewcommand{\csch}[2][says]{}

\title{ILP-based Local Search for Graph Partitioning}
\titlerunning{ILP-based Local Search for Graph Partitioning}

\author[1]{Alexandra Henzinger}
\author[2]{Alexander Noe}
\author[3]{Christian Schulz}
\affil[1]{Stanford University, Stanford, CA, USA\\ \texttt{ahenz@stanford.edu}}
\affil[2]{University of Vienna, Vienna, Austria\\
  \texttt{alexander.noe@univie.ac.at}}
\affil[3]{University of Vienna, Vienna, Austria\\ \texttt{christian.schulz@univie.ac.at}}
\authorrunning{A. Henzinger, A. Noe and C. Schulz}

\Copyright{Alexandra Henzinger, Alexander Noe, Christian Schulz}

\subjclass{G.2.2 Graph Theory}
\keywords{Graph Partitioning, Integer Linear Programming}

\date{today}

\newcommand{\CC}{C\texttt{++}}

\pgfplotscreateplotcyclelist{black white 2}{%
	every mark/.append style={solid},mark=*\\%
	every mark/.append style={solid},mark=square*\\%
	every mark/.append style={solid},mark=o\\%
	mark=star\\%
	every mark/.append style={solid},mark=diamond*\\%
	densely dashed,every mark/.append style={solid},mark=*\\%
	densely dashed,every mark/.append style={solid},mark=square*\\%
	densely dashed,every mark/.append style={solid},mark=o\\%
	densely dashed,every mark/.append style={solid},mark=star\\%
	densely dashed,every mark/.append style={solid},mark=diamond*\\%
}

\pgfplotsset{
  cycle list/Dark2,
  every axis/.append style={
    ylabel near ticks,
    log basis y={2},
    log basis x={2},
    legend cell align={left},
    legend style={font=\Large},
    label style={font=\Large},
    title style={font=\Large},
    tick label style={font=\Large},
    cycle multiindex* list={
      Dark2
      \nextlist
      black white 2
      \nextlist
   },
  }
}

\begin{document}
 \maketitle
 \begin{abstract}
Computing high-quality graph partitions is a challenging problem with numerous applications.  In this paper, we present a novel meta-heuristic for the balanced graph partitioning problem.  Our approach is based on integer linear programs that solve the partitioning problem to optimality.  However, since those programs typically do not scale to large inputs, we adapt them to heuristically improve a given partition.  We do so by defining a much smaller model that allows us to use symmetry breaking and other techniques that make the approach scalable. For example, in Walshaw’s well-known benchmark tables we are able to improve roughly half of all entries when the number of blocks is high.
\end{abstract}

\section{Introduction}
\emph{Balanced graph partitioning} is an important problem in computer science and engineering with an abundant amount of application domains, such as VLSI circuit design, data mining and distributed systems~\cite{gpApplicationPaper}.
It is well known that this problem is NP-complete~\cite{BuiJ92} and that no approximation algorithm with a constant ratio factor exists for general graphs unless P=NP~\cite{BuiJ92}.
Still, there is a large amount of literature on methods (with worst-case exponential time) that solve the graph partitioning problem to optimality.
This includes methods dedicated to the bipartitioning case \cite{armbruster2007branch,Armbruster2008, delling2012exact, delling2012better,feldmann2011n,felner2005,karisch2000solving,HagerPZ13,gp:lp,sellmann2003multicommodity} and some methods that solve the general graph partitioning problem~\cite{ferreira1998node,sensen2001lower}.
Most of these methods rely on the branch-and-bound framework \cite{land1960automatic}.
However, these methods can typically solve only very small problems as their running time grows exponentially, or if they can solve large bipartitioning instances using a moderate amount of time \cite{delling2012exact, delling2012better}, the running time highly depends on the bisection width of the graph.
Methods that solve the general graph partitioning problem \cite{ferreira1998node,sensen2001lower} have huge running times for graphs with up to a few hundred vertices.
Thus in practice mostly heuristic algorithms are used.

Typically the graph partitioning problem asks for a partition of a graph into $k$ blocks of about equal size such that there are few edges between them. 
Here, we focus on the case when the bounds on the size are very strict, including the case of \emph{perfect balance} when the maximal block size has to equal the average block size.

Our focus in this paper is on solution quality, \ie minimize the number of edges that run between blocks.
During the past two decades there have been numerous researchers trying to improve the best graph partitions in Walshaw's well-known partitioning benchmark~\cite{soper2004combined,wswebsite}.
Overall there have been more than forty different approaches that participated in this benchmark.
Indeed, high solution quality is of major importance in applications such as VLSI Design \cite{alpert1995rdn,alpert1999spectral} where even minor improvements in the objective can have a large impact on the production costs and quality of a chip. 
High-quality solutions are also favorable in applications where the graph needs to be partitioned only once and then the partition is used over and over again, implying that the running time of the graph partitioning algorithms is of a minor concern~\cite{DellingGPW11,heuvelinecoop,klsv-dtdch-10,Lau04,wagner2005pgs,ls-csarr-12}. Thirdly, high-quality solutions are even important in areas in which the running time overhead is paramount \cite{soper2004combined}, such as finite  element computations \cite{schloegel2000gph} or the direct solution of sparse linear systems \cite{george1973nested}.
Here, high-quality graph partitions can be useful for benchmarking purposes, \ie measuring how much more running time can be saved by higher quality solutions.

In order to compute high-quality solutions, state-of-the-art local search algorithms exchange vertices between blocks of the partition trying to decrease the cut size while also maintaining balance. This highly restricts the set of possible improvements.
Recently, we introduced new techniques that relax the balance constraint for vertex movements but globally maintain balance by combining multiple local searches~\cite{kabapeE}.  
This was done by reducing this combination problem to finding negative cycles in a graph.
In this paper, we extend the neighborhood of the combination problem by employing integer linear programming.
This enables us to find even more complex combinations and hence to further improve solutions.
More precisely, our approach is based on integer linear programs that solve the partitioning problem to optimality. 
However, out of the box those programs typically do not scale to large inputs, in particular because the graph partitioning problem has a very large amount of symmetry -- given a partition of the graph, each permutation of the block IDs gives a solution having the same objective and balance. 
Hence, we adapt the integer linear program to improve a given input partition. 
We do so by defining a much smaller graph, called \emph{model}, and solve the graph partitioning problem on the model to optimality by the integer linear program. More specifically, we select vertices close to the cut of the given input partition for potential movement and contract all remaining vertices of a block into a single vertex. A feasible partition of this model corresponds to a partition of the input graph having the same balance and objective.
Moreover, this model enables us to use symmetry breaking, which allows us to scale to much larger inputs.
To make the approach even faster, we combine it with initial bounds on the objective provided by the input partition, as well as providing the input partition to the integer linear program solver.
Overall, we arrive at a system that is able to improve more than half of all entries in Walshaw's benchmark when the number of blocks is high.

The rest of the  paper is organized as follows.  
We begin in Section~\ref{s:preliminaries} by introducing basic concepts. 
After presenting some related work in Section~\ref{s:related}
we outline the integer linear program as well as our novel local search algorithm in Section~\ref{s:algorithm}. 
Here, we start by explaining the very basic idea that allows us to find combinations of simple vertex movements. 
We then explain our strategies to improve the running time of the solver and strategies to select vertices for movement.
A summary of extensive experiments done to evaluate the performance of our algorithms is presented in Section~\ref{s:experiments}.
Finally, we conclude in Section~\ref{s:conclusion}.

\section{Preliminaries}

\label{s:preliminaries}
\subsection{Basic concepts}
Let $G=(V=\{0,\ldots, n-1\},E)$ be an undirected graph.  We
  consider positive, real-valued edge and vertex weight functions $\omega$ resp. $c$
  and extend them to sets, i.e., $\omega(E') \Is \sum_{x\in E'}\omega(x)$ and $c(V')\Is \sum_{x\in V'} c(x)$. Let $N(v)\Is
  \setGilt{u}{\set{v,u}\in E}$ denote the neighbors of $v$.  The
  degree of a vertex $v$ is $d(v):=|N(v)|$.  A vertex is a \emph{boundary
    vertex} if it is incident to at least one vertex in a different block.  We
  are looking for disjoint \emph{blocks} of vertices
  $V_1$,\ldots,$V_k$ that partition $V$; i.e., $V_1\cup\cdots\cup
  V_k=V$. The \emph{balancing constraint} demands that each block
has weight $c(V_i)\leq (1+\epsilon)\lceil\frac{c(V)}{k}\rceil=:L_{\max}$
for some imbalance parameter
$\epsilon$.      We call a block~$V_i$ \emph{overloaded} if its weight exceeds $L_{\max}$.
The objective of the problem is to minimize the total
  \emph{cut} $\omega(E\cap\bigcup_{i<j}V_i\times V_j)$ subject to the balancing constraints.
  We define the \emph{gain} of a vertex as the maximum decrease in the cut value when moving it to a different block.

\section{Related Work}
\label{s:related}
    There has been a \emph{huge} amount of research on graph partitioning and we refer the reader to the surveys given in~\cite{GPOverviewBook,SPPGPOverviewPaper,schloegel2000gph,Walshaw07} for most of the material.
    Here, we focus on issues closely related to our main contributions.
    All general-purpose methods that are able to obtain good partitions for large real-world graphs are based on the multi-level principle.
    Well-known software packages based on this approach include Jostle~\cite{Walshaw07}, KaHIP~\cite{kaffpa},~Metis~\cite{karypis1998fast}~and~Scotch~\cite{scotch}.

    Chris Walshaw's well-known benchmark archive has been established in 2001~\cite{soper2004combined,wswebsite}. Overall it contains 816 instances (34 graphs, 4 values of imbalance, and 6 values of $k$).
    Ever since there have been more than forty different approaches that participated in this benchmark.
    In this benchmark, the running time of the participating algorithms is not measured or reported.
Submitted partitions will be validated and added to the archive if they improve on a particular result.
This can either be an improvement in the number of cut edges or, if they match the current best cut size, an improvement in the weight of the largest block.
Most entries in the benchmark have as of Feb. $2018$ been obtained by Galinier~\etal\cite{galinier2011efficient} (more precisely an implementation of that approach by Frank Schneider), Hein and Seitzer~\cite{hein2011beyond} and the Karlsruhe High-Quality Graph Partitioning (KaHIP) framework~\cite{kabapeE}.
More precisely, Galinier \etal\cite{galinier2011efficient} use a memetic algorithm that is combined with tabu search to compute solutions and Hein and Seitzer~\cite{hein2011beyond} solve the graph partitioning problem by providing tight relaxations of a semi-definite program into a continuous problem.

The Karlsruhe High-Quality Graph Partitioning (\emph{KaHIP}) framework implements many different algorithms, for example flow-based methods and more-localized local searches, as well as several coarse-grained parallel and sequential meta-heuristics.
KaBaPE~\cite{kabapeE} is a coarse-grained parallel evolutionary algorithm, \ie each processor has its own population (set of partitions) and a copy of the graph.
After initially creating the local population, each processor performs multi-level combine and mutation operations on the local population. This is combined with a meta-heuristic that combines local searches that individually violate the balance
constraint into a more global feasible improvement.
For more details, we refer the reader to \cite{kabapeE}.

\section{Local Search based on Integer Linear Programming}
\label{s:algorithm}
We now explain our algorithm that combines integer linear programming and local search. We start by explaining the integer linear program that can solve the graph partitioning problem to optimality.
However, out-of-the-box this program does not scale to large inputs, in particular because the graph partitioning problem has a very large amount of symmetry. Thus, we reduce the size of the graph by first computing a partition using an existing heuristic and based on it collapsing parts of the graph. Roughly speaking, we compute a small graph, called \emph{model}, in which we only keep a small amount of selected vertices for potential movement and perform graph contractions on the remaining ones. A partition of the model corresponds to a partition of the input network having the same objective and balance. The computed model is then solved to optimality using the integer linear program. As we will see this process enables us to use symmetry breaking in the linear program, which in turn drastically speeds up computation times.

\subsection{Integer Linear Program for the Graph Partitioning Problem}
\label{ss:ilp}

We now introduce a generalization of an integer linear program formulation for balanced bipartitioning~\cite{brillout2009multi} to the general graph partitioning problem.
First, we introduce binary decision variables for all edges and vertices of the graph. More precisely, for each edge $e=\{u,v\}\in E$, we introduce the variable $e_{uv} \in \{0,1\}$ which is one if $e$ is a cut edge and zero otherwise. Moreover, for each $v\in V$ and block $k$, we introduce the variable $x_{v,k} \in \{0,1\}$ which is one if $v$ is in block $k$ and zero otherwise. Hence, we have a total of $|E| + k|V|$ variables.
We use the following constraints to ensure that the result is a valid $k$-partition:
\begin{ceqn}
  \begin{align}
  \label{eq:ilp-con-1a}
 \forall \{u,v\} \in E, \forall k&: e_{uv} \geq x_{u,k} - x_{v,k}\\
  \label{eq:ilp-con-1b}
 \forall \{u,v\} \in E, \forall k&: e_{uv} \geq x_{v,k} - x_{u,k}\\
  \label{eq:ilp-con-2}
  \forall k&: \sum_{v \in V} x_{v,k} c(v) \leq L_{\text{max}}\\
  \label{eq:ilp-con-3}
    \forall v \in V&: \sum_k x_{v,k} = 1
  \end{align}
\end{ceqn}

The first two constraints ensure that $e_{uv}$ is set to one if the vertices $u$ and $v$ are in different blocks. For an edge $\{u,v\}\in E$ and a block $k$, the right-hand side in this equation is one if one of the vertices $u$ and $v$ is in block $k$ and the other one is not. If both vertices are in the same block then the right-hand side is zero for all values of $k$. Hence, the variable can either be zero or one in this case. However, since the variable participates in the objective function and the problem is a minimization problem, it will be zero in an optimum solution.
The third constraint ensures that the balance constraint is satisfied for each partition. And finally, the last constraint ensures that each vertex is assigned to exactly one block. To sum up, our program has $2k|E|+k+|V|$ constraints and  $k \cdot (6 |E| + 2 |V|)$ non-zeros. Since we want to minimize the weight of cut edges, the objective function of our program is~written~as:

\begin{ceqn}
\begin{equation}
  \label{eq:ilp-sum}
  \min \sum_{\{u,v\} \in E} e_{uv} \cdot \omega(\{u,v\})
\end{equation}
\end{ceqn}

\subsection{Local Search}
The graph partitioning problem has a large amount of symmetry -- each permutation of the block IDs gives a solution with equal objective and balance.
Hence, the integer linear program described above will scan many branches that contain essentially the same solutions so that the program does not scale to large instances. 
Moreover, it is not immediately clear how to improve the scalability of the program by using symmetry breaking or~other~techniques. 

Our goal in this section is to develop a local search algorithm using the integer linear program above. 
Given a partition as input to be improved, our \emph{main idea} is to contract vertices ``that are far away'' from the cut of the partition.
In other words, we want to keep vertices close to the cut and contract all remaining vertices into one vertex for each block of the input partition.
This ensures that a partition of the contracted graph yields a partition of the input graph with the same objective and balance.
Hence, we apply the integer linear program to the model and solve the partitioning problem on it to optimality. Note, however, that due to the performed contractions this does not imply an optimal solution on~the~input~graph.

\begin{figure}[t]
\centering
\includegraphics[width=4.5cm]{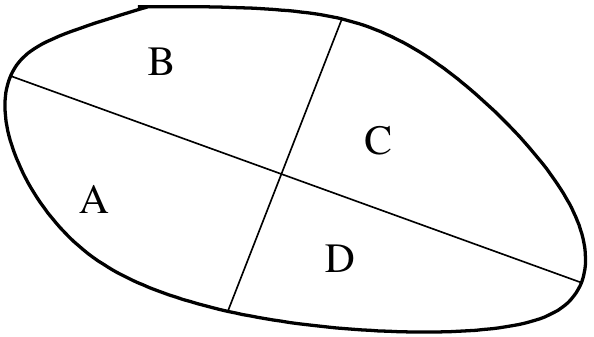}
\includegraphics[width=4.5cm]{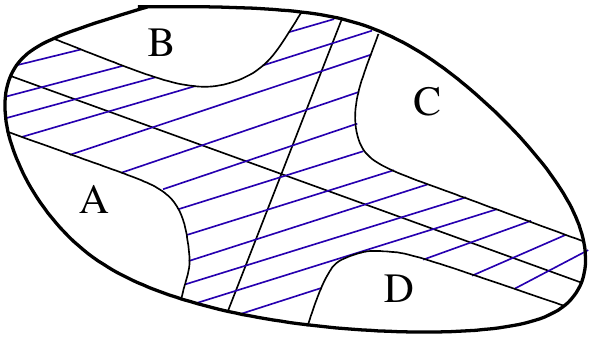}
\includegraphics[width=4.5cm]{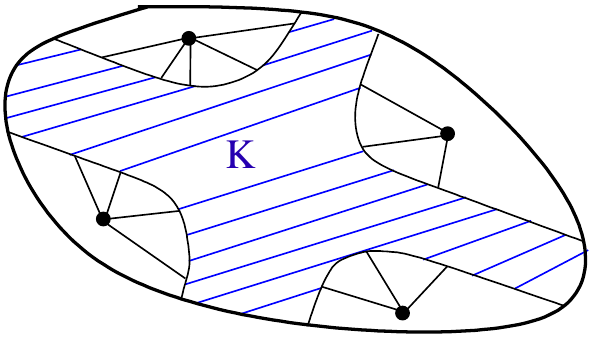}
\caption{From left to right: a graph that is partitioned into four blocks, the set $\mathcal{K}$ close to the boundary that will stay in the model, and lastly the model in which the sets $V_i \setminus \mathcal{K}$ have been~contracted.}
\label{fig:abstractexample}
\end{figure}

We now outline the details of the algorithm. Our local algorithm has two inputs, a graph~$G$ and a partition $V_1, \ldots, V_k$ of its vertices.
For now assume that we have a set of vertices $\mathcal{K} \subset V$ which we want to keep in the coarse model, \ie a set of vertices which we do not want to contract.
We outline in Section~\ref{ss:find} which strategies we have to select the vertices $\mathcal{K}$.
For the purpose of contraction we define $k$ sets $\mathcal{V}_i := V_i \setminus \mathcal{K}$. 
We obtain our coarse model by contracting each of these vertex sets. The contraction of a vertex set $\mathcal{V}_i$ works as follows:
the set of vertices is contracted into a single vertex $\mu_i$. 
The weight of $\mu_i$ is set to the sum of the weight of all vertices in the set that is contracted. 
There is an edge between two vertices $\mu_i$ and $v$ in the contracted graph if there
is an edge between a vertex of the set and $v$ in the original graph $G$.
The weight of an edge $(\mu_i,v)$ is set to the sum of the weight of edges that run between the vertices of~the~set~and~$v$. 
After all contractions have been performed the coarse model contains $k+|\mathcal{K}|$ vertices, and potentially much less edges than the input graph. Figure~\ref{fig:abstractexample} gives an abstract example of our model.

There are two things that are important to see: first, due to the way we perform contraction, the given partition of the input network yields a partition of our coarse model that has the same objective and balance simply by putting $\mu_i$ into block $i$ and keeping the block of the input for the vertices in $\mathcal{K}$. Moreover, if we compute a new partition of our coarse model, we can build a partition in the original graph with the same properties by putting the vertices~$\mathcal{V}_i$ into the block of their coarse representative $\mu_i$ together with the vertices of $\mathcal{K}$ that are in this block. Hence, we can solve the integer linear program on the coarse model to compute a~partition~for~the~input~graph.
After the solver terminates, \ie found an optimum solution of our mode or has reached a predefined time limit $\mathcal{T}$, we transfer the best solution to the original graph. Note that the latter is possible since an integer linear program solver typically computes intermediate solutions that may not be optimal.

\subsection{Optimizations}
\label{ss:opti}
Independent of the vertices $\mathcal{K}$ that are selected to be kept in the coarse model, the approach above allows us to define optimizations to solve our integer linear program faster.
We apply four strategies: (i) symmetry breaking, (ii) providing a start solution to the solver, (iii) add the objective of the input as a constraint as well as (iv) using the parallel solving facilities of the underlying solver. We outline the first three strategies in greater detail:

\subparagraph*{Symmetry Breaking.}
If the set $\mathcal{K}$ is small, then the solver will find a solution much faster. Ideally, our algorithms selects the vertices $\mathcal{K}$ such that $c(\mu_i)+c(\mu_j) > L_{\text{max}}$. In other words, no two contracted vertices can be clustered in one block. We can use this to break symmetry in our integer linear programming by adding constraints that fix the block of $\mu_i$ to block $i$, \ie we set $x_{\mu_i,i} = 1$ and $x_{\mu_i,j} =0$ for  $i\neq j$. Moreover, for those vertices we can remove the constraint which ensures that the vertex is assigned to a single unique block---since we assigned those vertices to a block using the new~additional~constraints.
\subparagraph*{Providing a Start Solution to the Solver.}
The integer linear program performs a significant amount of work in branches which correspond to solutions that are worse than the input partitioning.
Only very few - if any - solutions  are better than the given partition. However, we already know a fairly good partition (the given partition from the input) and give this partition to the solver by setting according initial values for all variables. This ensures that the integer linear program solver can omit many branches and hence speeds up the time needed to solve the integer linear program.

\subparagraph*{Solution Quality as a Constraint.}

Since we are only interested in improved partitions, we can add an additional constraint that disallows solutions which have a worse objective than the input partition. Indeed, the objective function of the linear program is linear, and hence the additional constraint is also linear.
Depending on the objective value, this reduces the number of branches that the linear program solver needs to look at. However, note that this comes at the cost of an additional constraint that needs to be evaluated.

\subsection{Vertex Selection Strategies}
\label{ss:find}
The algorithm above works for different vertex sets $\mathcal{K}$ that should be kept in the coarse model. There is an obvious trade-off: on the one hand, the set $\mathcal{K}$ should not be too large, otherwise the coarse model would be large and hence the linear programming solver needs a large amount of time to find a solution. On the other hand, the set should also not be too small, since this restricts the amount of possible vertex movements, and hence the approach is unlikely to find an improved solution. We now explain different strategies to select the vertex set $\mathcal{K}$. In any case, while we add vertices to the set $\mathcal{K}$, we compute the number of non-zeros in the corresponding ILP. We stop to add vertices when the number of non-zeros in the corresponding ILP is larger than a parameter $\mathcal{N}$.

\subparagraph*{Vertices Close to Input Cut.}
The intuition of the first strategy, \texttt{Boundary}, is that changes or improvements of the partition will occur reasonable close to the input partition. 
In this simple strategy our algorithm tries to use all \emph{boundary vertices} as the set $\mathcal{K}$.
In order to adhere to the constraint on the number of non-zeros in the ILP, we add the vertices of the boundary uniformly at random and stop if the number of non-zeros $\mathcal{N}$ is reached. 
If the algorithm managed to add all boundary vertices whilst not exceeding the specified number of non-zeros, we do the following extension: 
we perform a breadth-first search that is initialized with a random permutation of the boundary vertices. All additional vertices that are reached by the BFS are added to $\mathcal{K}$. As soon as the number of non-zeros $\mathcal{N}$ is reached, the algorithm~stops. \\

\subparagraph*{Start at Promising Vertices.}
Especially for high values of $k$ the boundary contains many vertices. The \texttt{Boundary} strategy quickly adds a lot of random vertices while ignoring vertices that have high gain.
However, note that even in good partitions it is possible that vertices with positive gain exist but cannot be moved due to the balance constraint.

Hence, our second strategy, \texttt{Gain}$_\rho$, tries to fix this issue by starting a breadth-first search initialized with only high gain vertices.
More precisely, we initialize the BFS with each vertex having gain $\geq \rho$ where~$\rho$~is~a~tuning~parameter.
Our last strategy, \texttt{TopVertices}$_\delta$, starts by sorting the boundary vertices by their gain.
We break ties uniformly at random.
Vertices are then traversed in decreasing order (highest gain vertices first) and for each start vertex $v$ our algorithm adds all vertices with distance $\leq \delta$ to the model.
The algorithm stops as soon as the number of non-zeros exceeds $\mathcal{N}$.

Early gain-based local search heuristics for the $\epsilon$-balanced graph partitioning problem searched for pairwise swaps with positive gain~\cite{fiduccia1982lth,Kernighan70}. More recent algorithms generalized this idea to also search for cycles or paths with positive total gain~\cite{kabapeE}. An important advantage of our new approach is that we solve the combination problem to optimality, \ie our algorithm finds the best combination of vertex movements of the vertices in $\mathcal{K}$ w.r.t to the input partition of the original graph. Therefore we can also find more complex optimizations that cannot be reduced to positive gain cycles and paths.

\vspace*{-.25cm}
\section{Experiments}\label{s:experiments}

\subsection{Experimental Setup and Methodology}

We implemented the algorithms using \CC\texttt{-17} and compiled all
codes using \texttt{g++-7.2.0} with full optimization (\texttt{-O3}). We use Gurobi 7.5.2 as an ILP solver and always use its parallel version. We perform experiments on the Phase 2 Haswell nodes of the SuperMUC supercomputer. The Phase 2 of SuperMUC consists of 3072 nodes, each with two Haswell Xeon E5-2697 v3 processors. Each node has 28 cores at 2.6GHz, as well as 64GB of main memory and runs the SUSE Linux Enterprise Server (SLES) operating system. Unless otherwise mentioned, our approach uses the shared-memory parallel variant of Gurobi using all 28 cores of a single node of the machine. 
In general, we perform five repetitions per instance and report the average running time as well as cut. Unless otherwise mentioned, we use a time limit for the integer linear program. When the time limit is passed, the integer linear program solver outputs the best solution that has currently been discovered. This solution does not have to be optimal.
Note that we do not perform experiments with Metis~\cite{karypis1998fast} and Scotch~\cite{scotch} in here, since previous papers, \eg \cite{kaffpa,kaffpaE}, have already shown that solution quality obtained is much worse than results achieved in the Walshaw benchmark.
When averaging over multiple instances, we use the geometric mean in order to give every instance the same influence on the \textit{final score}.  

\subparagraph*{Performance Plots.}
These plots relate the fastest running time to the running time of each other ILP-based local search algorithm on a per-instance basis.
For each algorithm, these ratios are sorted in increasing order. The plots show the ratio $t_\text{best}/t_\text{algorithm}$ on the y-axis to highlight the instances in which each algorithm performs badly. For plots in which we measure solution quality, the y-axis shows the ratio cut$_\text{best}/$cut$_\text{algorithm}$.
A point close to zero indicates that the running time/quality of the algorithm
was considerably worse than the fastest/best algorithm on the same instance. A value of one therefore indicates that the corresponding algorithm was one of the fastest/best algorithms to compute the solution.
Thus an algorithm is considered to outperform another algorithm if its corresponding ratio values are above those of the other algorithm.
In order to include instances that hit the time limit, we set the corresponding values below \emph{zero} for ratio computations.

\subparagraph*{Instances.}
We perform experiments on two sets of instances. Set $A$ is used to determine the performance of the integer linear programming optimizations and to tune the algorithm.
We obtained these instances from the Florida Sparse Matrix collection \cite{UFsparsematrixcollection} and the~10th~DIMACS Implementation Challenge~\cite{benchmarksfornetworksanalysis} to test our algorithm.
Set $B$ are all graphs from Chris Walshaw's graph partitioning benchmark archive~\cite{soper2004combined,wswebsite}. This archive is a collection of instances from finite-element applications, VLSI design and is one of the default benchmarking sets~for~graph~partitioning.

Table~\ref{tab:test_instances_walshaw} gives basic properties of the graphs from both benchmark sets.
We ran the unoptimized integer linear program that solves the graph partitioning problem to optimality from Section~\ref{ss:ilp} on the five smallest instances from the Walshaw benchmark set. With a time limit of $30$ minutes, the solver has only been able to compute a solution for two graphs with $k=2$. For higher values of $k$ the solver was unable to find any solution in the time limit. 
Even applying feasible optimizations does not increase the amount of ILPs solved. 
Hence, we omit further experiments in which we run an ILP solver on the full graph.

\begin{table}[t]
  \centering
  \caption{Basic properties of the our benchmark instances. }
 \small
  \begin{tabular}{| l | r | r || l | r | r |}
    \hline
    Graph & $n$& $m$  & Graph & $n$ & $m$ \\
    \hline \hline
    \multicolumn{3}{|c||}{Walshaw Graphs (Set B)} &  \multicolumn{3}{c|}{Walshaw Graphs (Set B)}\\
    \hline

    add20       & \numprint{2395} & \numprint{7462} &wing        & \numprint{62032} & $\approx121$K \\
    data        & \numprint{2851} & \numprint{15093} &brack2      & \numprint{62631} & $\approx366$K\\
    3elt        & \numprint{4720} & \numprint{13722} &finan512    & \numprint{74752} & $\approx261$K\\
    uk          & \numprint{4824} & \numprint{6837} &fe\_tooth   & \numprint{78136} & $\approx452$K\\
    add32       & \numprint{4960} & \numprint{9462} &fe\_rotor   & \numprint{99617} & $\approx662$K \\
    bcsstk33    & \numprint{8738} & $\approx291$K &598a        & \numprint{110971} & $\approx741$K \\
    whitaker3   & \numprint{9800} & \numprint{28989} &fe\_ocean   & \numprint{143437} & $\approx409$K\\
    crack       & \numprint{10240}& \numprint{30380} &144         & \numprint{144649} & $\approx1.1$M\\
    wing\_nodal & \numprint{10937}& \numprint{75488} &wave        & \numprint{156317} & $\approx1.1$M\\
    fe\_4elt2   & \numprint{11143}& \numprint{32818} &m14b        & \numprint{214765} & $\approx1.7$M\\
    vibrobox    & \numprint{12328}& $\approx165$K &auto        & \numprint{448695} & $\approx3.3$M \\
    \cline{4-6}

    bcsstk29    & \numprint{13992}& $\approx302$K &\multicolumn{3}{c|}{}\\
    4elt        & \numprint{15606}& \numprint{45878} & \multicolumn{3}{c|}{Parameter Tuning (Set A)}\\
    \cline{4-6}
    fe\_sphere  & \numprint{16386}& \numprint{49152} & delaunay\_n15 & \numprint{32768} & \numprint{98274}\\
    cti         & \numprint{16840}& \numprint{48232} & rgg\_15 & \numprint{32768} & $\approx160$K\\
    memplus     & \numprint{17758}& \numprint{54196} & 2cubes\_sphere & \numprint{101492} & $\approx772$K\\
    cs4         & \numprint{22499}& \numprint{43858} & cfd2            & \numprint{123440} & $\approx1.5$M \\
    bcsstk30    & \numprint{28924}& $\approx 1.0$M & boneS01         & \numprint{127224} & $\approx3.3$M\\
    bcsstk31    & \numprint{35588}& $\approx572$K & Dubcova3        & \numprint{146689} & $\approx1.7$M\\
    fe\_pwt     & \numprint{36519}& $\approx144$K & G2\_circuit   & \numprint{150102} & $\approx288$K \\
    bcsstk32    & \numprint{44609}& $\approx985$K & thermal2  & \numprint{1227087} & $\approx3.7$M\\
    fe\_body    & \numprint{45087}& $\approx163$K & as365     & \numprint{3799275} & $\approx11.4$M\\
    t60k        & \numprint{60005}& \numprint{89440} & adaptive  & \numprint{6815744} & $\approx13.6$M\\

                                    \hline
  \end{tabular}
  \label{tab:test_instances_walshaw}
\end{table}

\subsection{Impact of Optimizations}
We now evaluate the impact of the optimization strategies for the ILP that we presented in Section~\ref{ss:opti}.  
In this section, we use the variant of our local search algorithm in which $\mathcal{K}$ is obtained by starting depth-one breadth-first search at the $25$ highest gain vertices, and set the limit on the non-zeros in the ILP to $\mathcal{N}=\infty$. However, we expect the results in terms of speedup to be similar for different vertex selection strategies.
To evaluate the ILP performance, we run KaFFPa using the strong preconfiguration on each of the graphs from set $A$ using $\epsilon=0$ and $k\in\{2,4,8,16,32,64\}$ and then use the computed partition as input to each ILP (with the different optimizations).
As the optimizations do not change the objective value achieved in the ILP, we only report running times of our different approaches. 
We set the time limit of the ILP solver to 30 minutes.

We use five variants of our algorithm: \texttt{Basic} does not contain any optimizations; \texttt{BasicSym} enables symmetry breaking; \texttt{BasicSymSSol} additionally gives the input partitioning to the ILP solver. The two variants \texttt{BSSSConst=} and \texttt{BSSSConst$<$} are the same as \texttt{BasicSymSSol} with additional constraints: \texttt{BSSSConst=} has the additional constraint that the objective has to be smaller or equal to the start solution,  \texttt{BSSSConst$<$} has the constraint that the objective value of a solution must be better than the objective value of the start solution. Figure~\ref{fig:plotresults} summarises the results.

\begin{figure}
  \centering
  \raisebox{2.2mm}{
    \includegraphics[width=6.5cm]{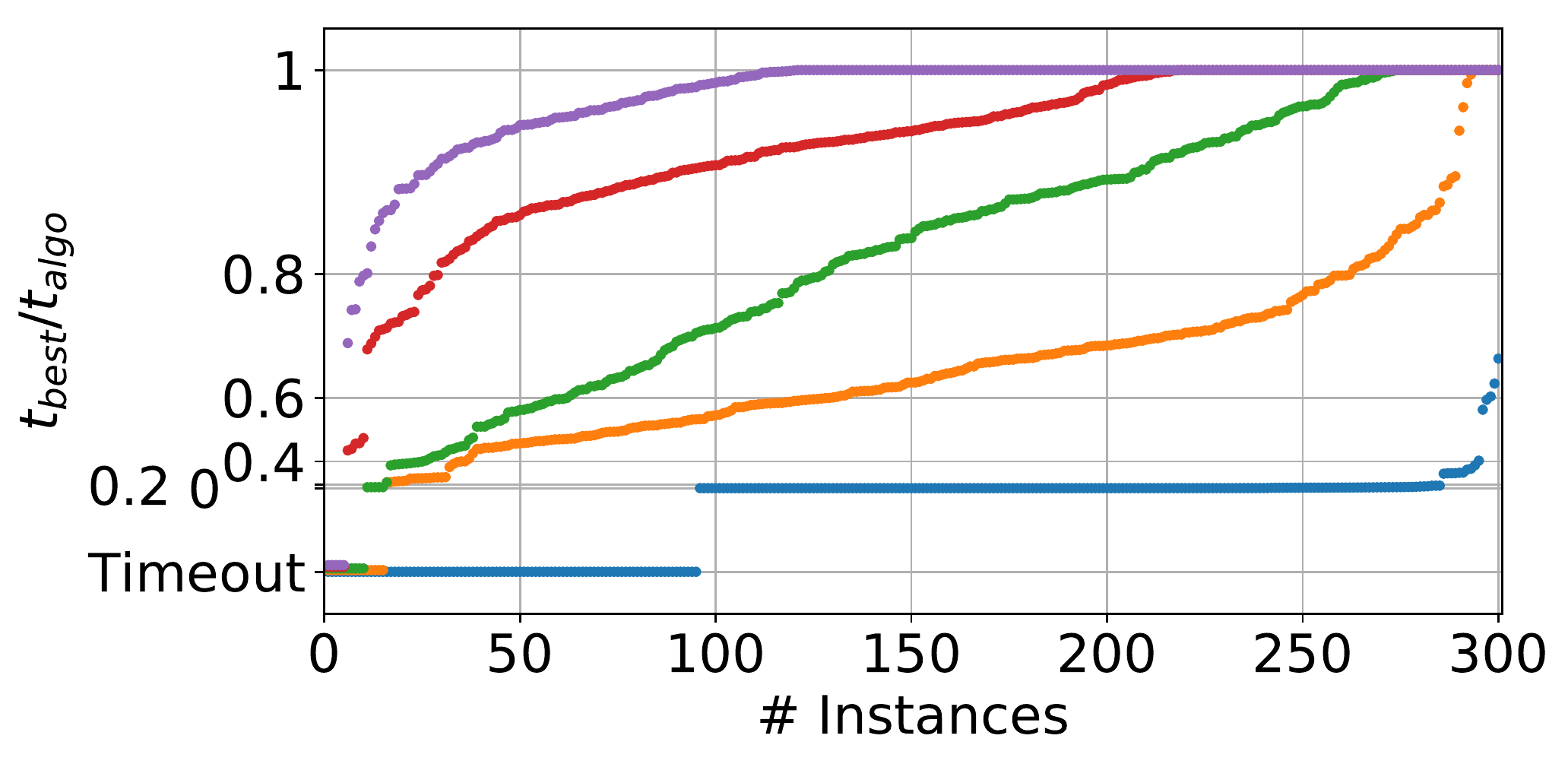}
    }
\includegraphics[width=7cm]{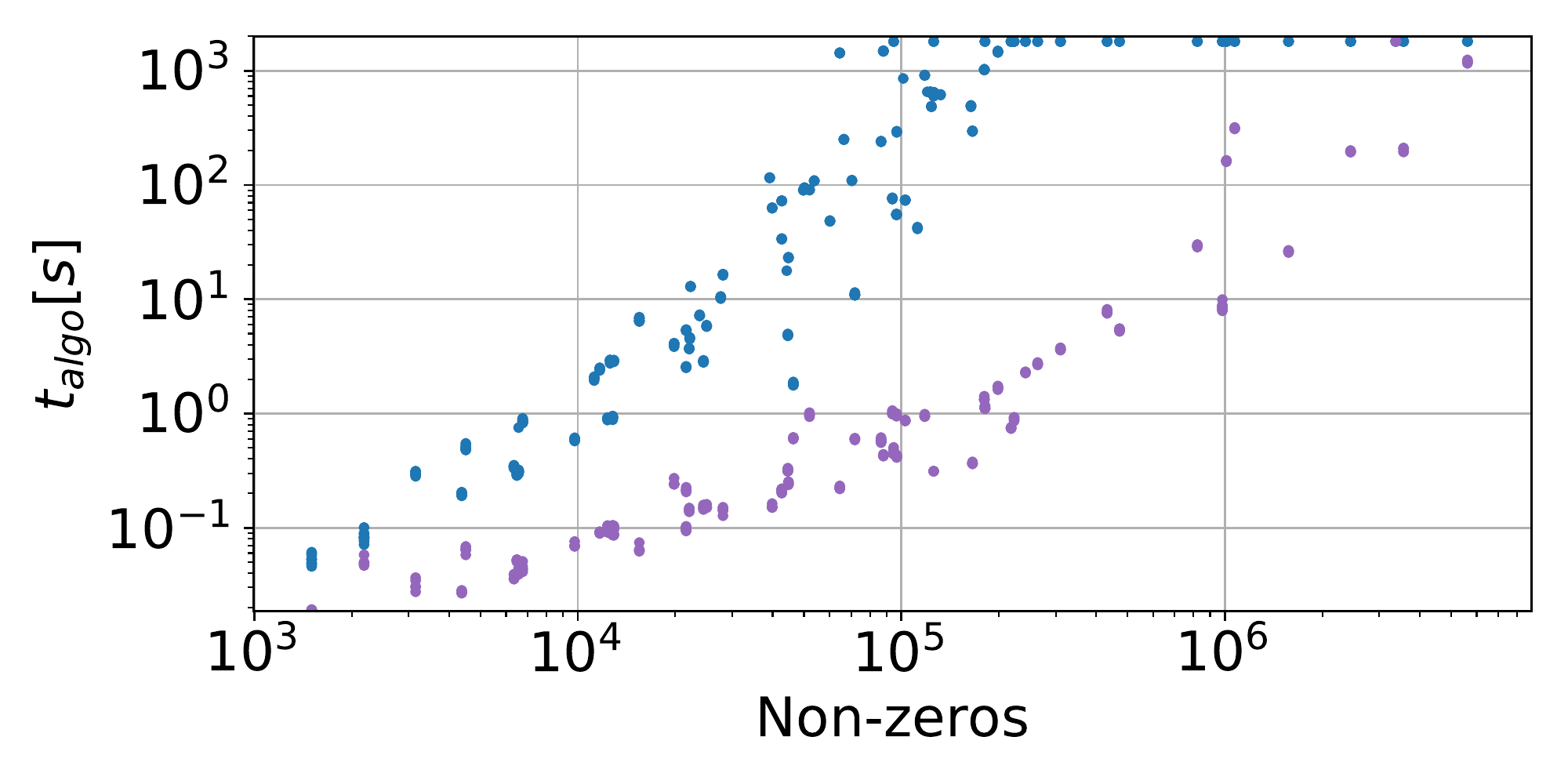}

\centering

  \includegraphics[width=11cm]{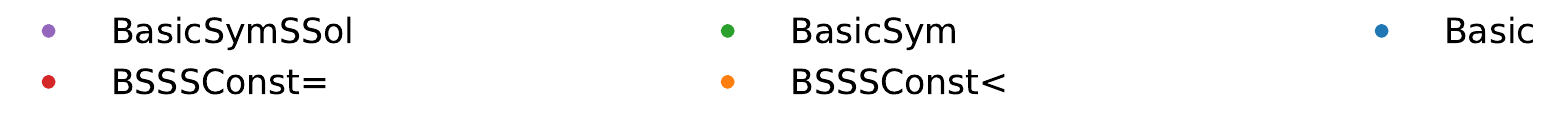}
\caption{
\emph{Left:} performance plot for five variants of our algorithm: \texttt{Basic} does not contain any optimizations; \texttt{BasicSym} enables symmetry breaking; \texttt{BasicSymSSol} additionally gives the input partitioning to the ILP solver. The two variants \texttt{BSSSConst=} and \texttt{BSSSConst$<$} are the same as \texttt{BasicSymSSol} with additional constraints: \texttt{BSSSConst=} has the additional constraint that the objective has to be smaller or equal to the start solution,  \texttt{BSSSConst$<$} has the constraint that the solution must be better than~the~start~solution. \emph{Right:} performance of the slowest (\texttt{Basic}) and fastest ILPs (\texttt{BasicSymSSol}) depending on the number of non-zeros in the ILP.
}
\label{fig:plotresultsilpvariants}
\end{figure}

In our experiments, the basic configuration reaches the time limit in 95 out of the 300 runs.
Overall, enabling symmetry breaking drastically speeds up computations.  On all of the instances which the \texttt{Basic} configuration could solve within the time limit, each other configuration is faster than the \texttt{Basic} configuration. Symmetry breaking speeds up computations by a factor of 41 in the geometric mean on those instances. The largest obtained speedup on those instances was a factor of 5663 on the graph adaptive for $k=32$. The configuration solves all but the two instances (boneS01, $k=32$) and (Dubcova3, $k=16$) within the time limit. Additionally providing the start solution (\texttt{BasicSymSSol}) gives an addition speedup of 22\% on average. Over the \texttt{Basic} configuration, the average speedup is 50 with the largest speedup being 6495 and the smallest speedup being 47\%. This configuration can solve all instances within the time limit except the instance boneS01 for $k=32$. Providing the objective function as a constraint (or strictly smaller constraint) does not further reduce the running time of the solver. Instead, the additional constraints even increase the running time. We adhere this to the fact that the solver has to do additional work to~evaluate~the~constraint.
We conclude that \texttt{BasicSymSSol} is the fastest configuration of the ILP. Hence, we use this configuration in all the following experiments. 
Moreover, from Figure~\ref{fig:plotresultsilpvariants} we can see that this configuration can solve most of the instance within the time limit if the number of non-zeros in the ILP is below $10^6$. Hence, we set the parameter $\mathcal{N}$ to~$10^6$ in the following section.

\vfill
\pagebreak
\subsection{Vertex Selection Rules}
\label{s:node_selection}

We now evaluate the vertex selection strategies to find the set of vertices $\mathcal{K}$ that model the ILP.  
We look at all strategies described in Section~\ref{ss:find}, \ie \texttt{Boundary}, \texttt{Gain$_{\rho}$} with the parameter $\rho \in \{-2,-1,0,1\}$ as well as \texttt{TopVertices}$_\delta$ for $\delta \in \{1,2,3\}$.
To evaluate the different selection strategies, we use the best of five runs of KaFFPa strong on each of the graphs from set $A$ using $\epsilon=0$ and $k\in\{2,4,8,16,32,64\}$ and then use the computed partition as input to the ILP (with~different~sets~$\mathcal{K}$). Table~\ref{table:variants} summarizes the results of the experiment, \ie the number of cases in which our algorithm was able to improve the result, the average running time in seconds for these selection strategies as well as the number of cases in which the strategy computed the best result (the partition having the lowest cut).
  We set the time limit to $2$ days to be able to finish almost all runs without running into timeout. For the average running time we exclude all graphs in which at least one algorithm did not finish in $2$ days (rgg\_15 $k=16$, delaunay\_n15 $k=4$, G2\_circuit $k=4,8$).
 If multiple runs share the best result, they are all counted. However, when no algorithm improves the input partition on a graph, we do not count them.

\begin{table}[b!]
    \centering
    \small

    \caption{From top to bottom: Number of improvements found by different vertex selection rules relative to the total number of instances, average running time of the strategy on the subset of instances (graph, $k$) in which all strategies finished within the time limit, and the relative number of instances in which the strategy computed the lowest cut. Best values are highlighted in bold. \label{table:variants}}
      \begin{tabular}{|l||rrr|rrr|r|}

        \hline
          &\texttt{Gain} &&&\texttt{TopVertices} &&& \texttt{Boundary} \\
      $k$& $\rho = 0 $& $\rho = -1 $ & $\rho=-2$ &$\delta = 1$ & $\delta=2$& $\delta=3$& \\\hline \hline
&      \multicolumn{7}{c|}{Relative Number of Improvements}  \\\hline
       2 & \textbf{70\%} & \textbf{70\%} & \textbf{70\%} & 50\% & \textbf{70\%} & \textbf{70\%} & \textbf{70\%} \\
       4 &          50\% &          60\% & \textbf{80\%} & 70\% &          70\% &          70\% & \textbf{80\%} \\
       8 &          50\% &          60\% & \textbf{78\%} & 60\% &          60\% &          60\% &          48\% \\
      16 &          30\% &          50\% & \textbf{70\%} & 40\% &          30\% &          30\% &          40\% \\
      32 & \textbf{60\%} & \textbf{60\%} &          46\% & 50\% &          50\% &          20\% &          20\% \\
      64 & \textbf{70\%} & \textbf{70\%} &          50\% & 30\% &          20\% &          20\% &           0\% \\
  \hline\hline
&      \multicolumn{7}{c|}{Average Running Time} \\
\hline
                2 & 189.943s & 292.573s & 357.145s &  \textbf{34.045s} &  61.152s &  92.452s &        684.198s \\
       4 & 996.934s & 628.950s & 428.353s &  \textbf{87.357s} & 255.223s & 558.578s &       1467.595s \\
       8 & 552.183s & 244.470s & 244.046s & 105.737s & 167.164s & 340.900s &         \textbf{96.763s} \\
      16 & 118.532s &  52.547s &  90.363s &  53.385s & 141.814s & 243.957s &         \textbf{34.790s} \\
               32 &  40.300s &  24.607s &  94.146s &  27.156s &  80.252s & 116.023s & \textbf{7.596s} \\
               64 &  15.866s &  21.908s &  24.253s &  14.627s &  30.558s &  44.813s & \textbf{4.187s} \\
      \hline
      \hline
&      \multicolumn{7}{c|}{Relative Number Best Algorithm} \\
\hline

   2 & 20\% & \textbf{60\%} &          50\% & 10\% & 10\% &  0\% & \textbf{60\%} \\
                4 & 10\% &           0\% & \textbf{50\%} & 10\% &  0\% &  0\% &          30\% \\
                8 &  0\% &          20\% & \textbf{30\%} & 10\% & 10\% & 10\% &          26\% \\
               16 &  0\% &          10\% & \textbf{54\%} & 10\% &  0\% & 10\% &          20\% \\
               32 &  0\% &           8\% & \textbf{38\%} &  0\% &  0\% &  0\% &           4\% \\
      64 &  0\% &          16\% &          \textbf{36\%} &  0\% &  0\% &  0\% &           0\% \\

        \hline
    \end{tabular}
  \end{table}

\begin{figure}
  \centering
  \raisebox{.0mm}{
    \includegraphics[width=7cm]{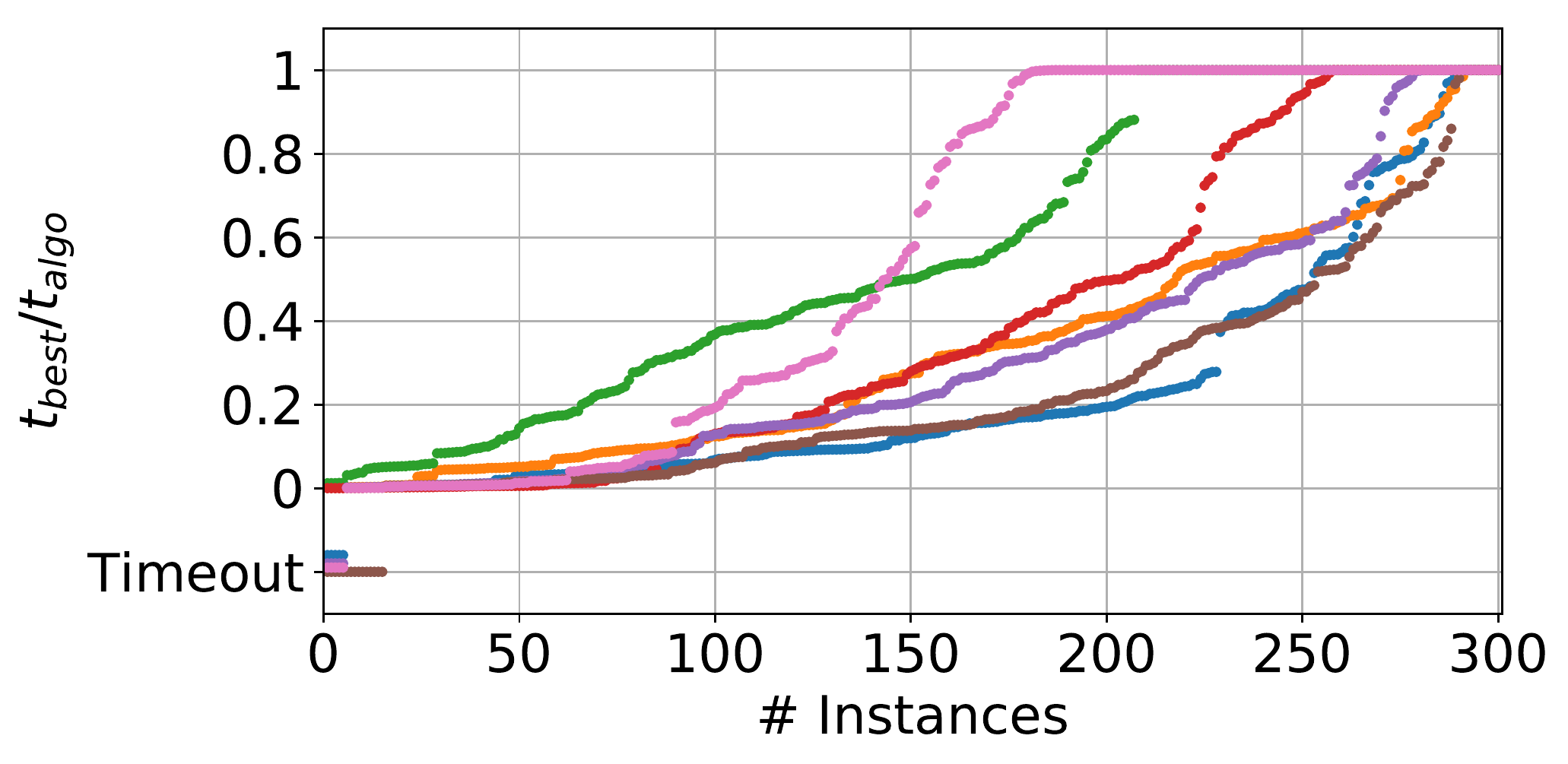}
    }
\includegraphics[width=6.5cm]{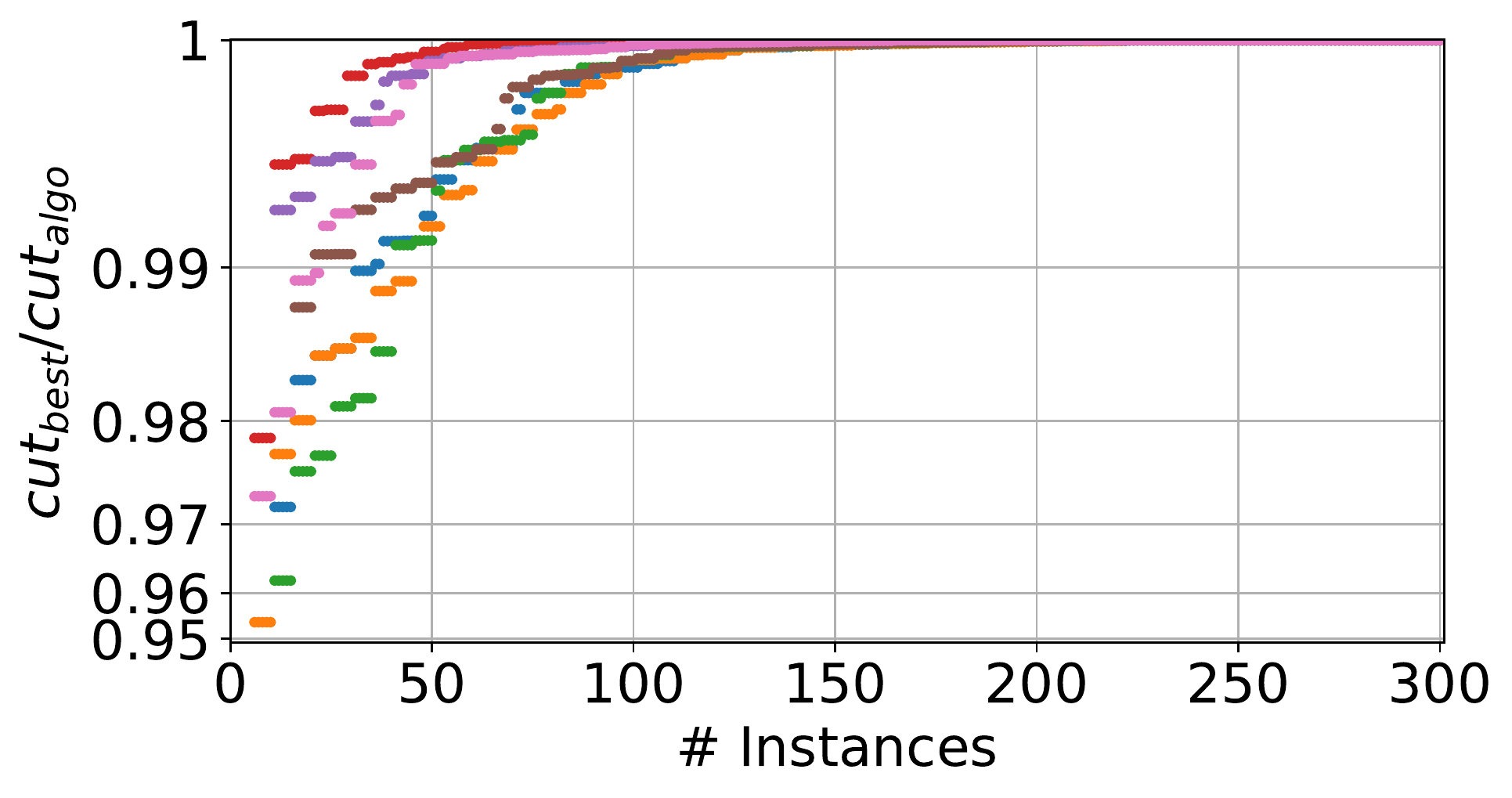}

\centering

  \includegraphics[width=11cm]{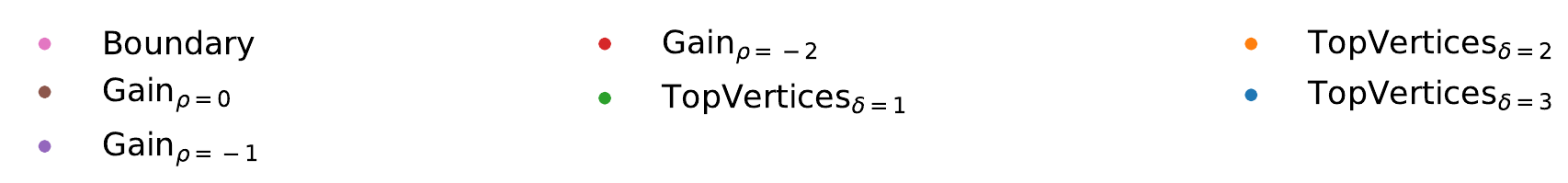}
\caption{
\emph{Left:} performance plot for all vertex selection strategies \emph{Right:} cut value of vertex selection strategies in comparison to the best result given by any strategy.
}
\label{fig:plotresults}
\vspace*{-.5cm}
\end{figure}

Looking at the number of improvements, the \texttt{Boundary} strategy is able to improve the input for small values of $k$, but with increasing number of blocks $k$ improvements decrease to no improvement in all runs with $k=64$.
Because of the limit on the number of non-zeros, the ILP contains only random boundary vertices for large values of $k$ in this case. 
Hence, there are not sufficiently many high gain vertices in the model and fewer improvements for large values of $k$ are expected.
For small values of $k \in \{2,4\}$, the \texttt{Boundary} strategy can improve as many as the \texttt{Gain$_{\rho=-2}$} strategy but the average running times are higher.

For $k = \{2,4,8,16\}$, the strategy \texttt{Gain$_{\rho=-2}$} has the highest number of improvements, for $k=\{32,64\}$ it is surpassed by the strategy \texttt{Gain$_{\rho=-1}$}. However, the strategy \texttt{Gain}$_{\rho=-2}$ finds the best cuts in most cases among all tested strategies. Due to the way these strategies are designed, they are able to put a lot of high gain vertices into the model as well as vertices that can be used to balance vertex movements. 
The \texttt{TopVertices} strategies are overall also able to find a large number of improvements. However, the found improvements are typically  smaller than the \texttt{Gain} strategies. 
This is due to the fact that the \texttt{TopVertices} strategies grow BFS balls with a predefined depth around high gain vertices first, and later on are not able to include vertices that could be used to balance their movement.
Hence, there are less potential vertex movements that could yield an improvement.

For almost all strategies, we can see that the average running time decreases as the number of blocks $k$ increases.
This happens because we limit the number of non-zeros $\mathcal{N}$ in our ILP. As the number of non-zeros grows linear with the underlying model size, the models are far smaller for higher values of $k$. 
Using symmetry breaking, we already fixed the block of the $k$ vertices $\mu_i$ which represent the vertices not part of $\mathcal{K}$. 
Thus the ILP solver can quickly prune branches which would place vertices connected heavily to one of these vertices in~a~different~block.
Additionally, our data indicate that a large number of small areas in our model results faster in solve times than when the model contains few large areas. The performance plot in Figure~\ref{fig:plotresults} shows that the strategies \texttt{Boundary}, \texttt{TopVertices}$_{\delta=1}$ and \texttt{Gain$_{\rho=-2}$} have lower running times than other strategies. These strategies all select a large number of vertices to initialize the breadth-first search. Therefore they output a vertex set $\mathcal{K}$ that is the union of many small areas around these vertices. Variants that initialize the breadth-first search with fewer vertices have fewer areas, however each of the areas is larger.

\subsection{Walshaw Benchmark}
In this section, we present the results when running our best configuration on all graphs from Walshaw's benchmark archive.
Note that the rules of the benchmark imply that running time is not an issue, but algorithms should achieve the smallest possible cut value while satisfying the balance constraint.
We run our algorithm in the following setting:
We take existing partitions from the archive and use those as input to our algorithm.
As indicated by the experiments in Section~\ref{s:node_selection}, the vertex selection strategies \texttt{Gain$_{\rho \in \{-1,-2\}}$} perform best for different values of $k$.
Thus we use the variant \texttt{Gain$_{\rho=-2}$} for $k \leq 16$ and both \texttt{Gain$_{\rho=-2}$} and \texttt{Gain$_{\rho=-1}$} otherwise in this section.
We repeat the experiment once for each
\begin{wraptable}{r}{0.45\textwidth}
\centering
        \caption{Relative number of improved instances in the Walshaw Benchmark starting from current entries reported in the Walshaw benchmark.}
 \begin{tabular}{|l|r|r|r|r|}
 $k$/$\epsilon$    & $0\%$ & $1\%$ & $3\%$ & $5\%$  \\
 \hline
 \hline
 2                 & $6\%$ & $12\%$ & $6\%$& $6\%$        \\
 4                 & $18\%$ &$9\%$& $6\%$&  $18\%$     \\
 8                 & $26\%$ &$24\%$& $12\%$& $15\%$     \\
 16                & $50\%$ &$26\%$& $29\%$& $29\%$     \\
 32                & $62\%$ &$47\%$& $47\%$& $53\%$    \\
 64                & $68\%$ &$59\%$& $71\%$& $76\%$   \\
 \hline
 \hline
   sum               & $38\%$ & $29\%$ & $28\%$ & $33\%$\\
         \end{tabular}
        \label{tab:existingimprovment}
      \end{wraptable}
      instance (graph, $k$)
      and run our algorithm for $k = \{2,4,8,16,32,64\}$ and $\epsilon \in \{0, 1\%, 3\%, 5\%\}$. For larger values of $k \in \{32,64\}$, we strengthen our strategy and use  $\mathcal{N} = 5 \cdot 10^6$ as a bound for the number of non-zeros.
Table~\ref{tab:existingimprovment} summarizes the results and Table~\ref{table:detailedimprovement} in the Appendix gives detailed per-instance results.

When running our algorithm using the currently best partitions provided in the benchmark, we are able to improve 38\% of the currently reported perfectly balanced results. 
We are able to improve a larger amount of results for larger values of $k$, more specifically, out of the partitions with $k \geq 16$, we can improve $60\%$ of all perfectly balanced partitions. This is due to the fact that the graph partitioning problem becomes more difficult for larger values of $k$.
There is a wide range of improvements with the smallest improvement being $0.0008\%$ for graph auto with $k=32$ and $\epsilon = 3\%$ and with the largest improvement that we found being $1.72\%$ for fe\_body for $k=32$ and $\epsilon = 0\%$. The largest absolute improvement we found is~$117$ for bcsstk32 with $k=64$ and $\epsilon = 0\%$. In general, the total number of improvements becomes less if more imbalance is allowed. This is also expected since traditional local search methods have a larger amount of freedom to move vertices. However, the number of improvements still shows that the method is also able to improve a large number of partitions for large values of $k$ even if more imbalance is allowed.
\vspace{-.25cm}
\section{Conclusions and Future Work}\label{s:conclusion}
We presented a novel meta-heuristic for the balanced graph partitioning problem. 
Our approach is based on an integer linear program that solves a model to combine unconstraint vertex movements into a global feasible improvement.
Through a given input partition, we were able to use symmetry breaking and other techniques that make the approach scale to large inputs.
In Walshaw’s well known benchmark tables, we were able to improve a large amount of partitions given in the benchmark.

In the future, we plan to further improve our implementation and integrate it into the KaHIP framework. We would like to look at other objective functions as long as they can be modelled linearly. Moreover, we want to investigate weather this kind of contractions can be useful for other ILPs. It may be interesting to find cores for contraction by using the information provided an evolutionary algorithm like KaFFPaE~\cite{kaffpaE}, \ie if many of the individuals of the population of the evolutionary algorithm agree that two vertices should be put together in a block then those should be contracted in our model. Lastly, besides using other exact techniques like branch-and-bound to solve our combination model, it may also be worthwhile to use a heuristic algorithm instead.

\vspace*{-.25cm}
\section*{Acknowledgements}
\vspace{-.25cm}
The research leading to these results has received funding from the European Research Council under the European Community's Seventh Framework Programme (FP7/2007-2013) /ERC grant agreement No. 340506.
Moverover, the  authors  gratefully  acknowledge  the  Gauss  Centre  for Supercomputing e.V.  (www.gauss-centre.eu) for funding this project by providing computing time on the GCS Supercomputer SuperMUC at Leibniz Supercomputing~Centre~(www.lrz.de).

\vfill\pagebreak
\renewcommand{\bibname}{\begin{flushleft} References \end{flushleft}}
\bibliographystyle{plainurl}
\bibliography{phdthesiscs,quellen}

\newpage
\begin{appendix}

\section{Additional Tables}

  \begin{table}[h!]
   \centering
    \small
      \caption{Improvement of existing partitions from the Walshaw benchmark with $\epsilon = 0\%$ using our ILP approach. In each $k$-column the results computed by our approach are on the left and the current Walshaw cuts are on the right. Results achieved by \texttt{Gain$_{\rho=-1}$} are marked with \^\ and results achieved by \texttt{Gain$_{\rho=-2}$} are marked with *. \label{table:detailedimprovement}}
\resizebox{\linewidth}{!}{

  \begin{tabular}{|l|rr|rr|rr|rr|rr|rr|}
\hline
      Graph / k & \multicolumn{2}{|c|}{2} & \multicolumn{2}{|c|}{4} & \multicolumn{2}{|c|}{8} & \multicolumn{2}{|c|}{16} & \multicolumn{2}{|c|}{32} & \multicolumn{2}{|c|}{64} \\ \hline \hline
            add20 &   596 &   596 &  1151 &  1151 &  1681 &  1681 &  2040 &  2040 &   \textbf{*2360} &   2361 &   \textbf{\^{}2947} &   2949 \\
             data &   189 &   189 &   382 &   382 &   668 &   668 &  1127 &  1127 &   1799 &   1799 &   2839 &   2839 \\
             3elt &    90 &    90 &   201 &   201 &   345 &   345 &   573 &   573 &    960 &    960 &   1532 &   1532 \\
               uk &    19 &    19 &    41 &    41 &    83 &    83 &   145 &   145 &    \textbf{*\^{}246} &    247 &    408 &    408 \\
            add32 &    11 &    11 &    34 &    34 &    67 &    67 &   118 &   118 &    213 &    213 &    485 &    485 \\
         bcsstk33 & 10171 & 10171 & 21717 & 21717 & 34437 & 34437 & 54680 & 54680 &  77414 &  77414 & 107185 & 107185 \\
        whitaker3 &   127 &   127 &   381 &   381 &   656 &   656 &  1085 &  1085 &   1668 &   1668 &   2491 &   2491 \\
            crack &   184 &   184 &   366 &   366 &   679 &   679 &  1088 &  1088 &   \textbf{*1678} &   1679 &   2535 &   2535 \\
      wing\_nodal &  1707 &  1707 &  3575 &  3575 &  5435 &  5435 &  \textbf{*8333} &  8334 &  11768 &  11768 &  \textbf{*\^{}15774} &  15775 \\
        fe\_4elt2 &   130 &   130 &   349 &   349 &   607 &   607 &  1007 &  1007 &   1614 &   1614 &   2475 &   2478 \\
         vibrobox & 10343 & 10343 & 18976 & 18976 & 24484 & 24484 & \textbf{*\^{}31848} & 31850 &  \textbf{*39474} &  39477 &  \textbf{*46568} &  46571 \\
         bcsstk29 &  2843 &  2843 &  8035 &  8035 & 13975 & 13975 & 21905 & 21905 &  \textbf{*34733} &  34737 &  55241 &  55241 \\
             4elt &   139 &   139 &   326 &   326 &   545 &   545 & \textbf{*\^{}933} &   934 &   1551 &   1551 &   \textbf{\^{}2564} &   2565 \\
       fe\_sphere &   386 &   386 &   768 &   768 &  1156 &  1156 &  1714 &  1714 &   2488 &   2488 &   3543 &   3543 \\
              cti &   334 &   334 &   954 &   954 &  1788 &  1788 &  2793 &  2793 &   4046 &   4046 &   5629 &   5629 \\
          memplus &  \textbf{*5499} &  5513 & \textbf{*9442} &  9448 & \textbf{*\^{}11710} & 11712 & \textbf{\^{}12893} & 12895 &  \textbf{*\^{}13947} &  13953 &  \textbf{\^{}16188} &  16223 \\
              cs4 &   369 &   369 &   932 &   932 &  1440 &  1440 &  2075 &  2075 &   \textbf{*2907} &   2928 &   \textbf{\^{}4025} &   4027 \\
         bcsstk30 &  6394 &  6394 & 16651 & 16651 & 34846 & 34846 & \textbf{*\^{}70407} & 70408 & 113336 & 113336 & \textbf{*171148} & 171153 \\
         bcsstk31 &  2762 &  2762 &  7351 &  7351 & \textbf{*13280} & 13283 & \textbf{*23857} & 23869 &  \textbf{*37143} &  37158 &  \textbf{*57354} &  57402 \\
          fe\_pwt &   340 &   340 &   705 &   705 &  1447 &  1447 &  2830 &  2830 &   \textbf{*\^{}5574} &   5575 &   \textbf{\^{}8177} &   8180 \\
         bcsstk32 &  4667 &  4667 &  9311 &  9311 & \textbf{*\^{}20008} & 20009 & \textbf{*\^{}36249} & 36250 &  \textbf{*60013} &  60038 &  \textbf{*90778} &  90895 \\
         fe\_body &   262 &   262 &   599 &   599 &  1033 &  1033 &  \textbf{*1722} &  1736 &   \textbf{\^{}2797} &   2846 &   \textbf{*4728} &   4730 \\
             t60k &    79 &    79 &   209 &   209 &   456 &   456 &   \textbf{\^{}812} &   813 &   1323 &   1323 &  \textbf{*\^{}2074} &   2077 \\
             wing &   789 &   789 &  1623 &  1623 &  2504 &  2504 &  \textbf{\^{}3870} &  3876 &   \textbf{\^{}5592} &   5594 &   \textbf{\^{}7622} &   7625 \\
           brack2 &   731 &   731 &  3084 &  3084 &  7140 &  7140 & 11570 & 11570 &  \textbf{\^{}17382} &  17387 &  \textbf{*25805} &  25808 \\
      finan512 &   162 &   162 &   324 &   324 &   648 &   648 &  1296 &  1296 &   2592 &   2592 &  10560 &  10560 \\
      fe\_tooth &  3816 &  3816 &  \textbf{*6888} &  6889 & \textbf{*11414} & 11418 & \textbf{*\^{}17352} & 17355 &  \textbf{*24879} &  24885 &  \textbf{*34234} &  34240 \\
        fe\_rotor &  2098 &  2098 &  7222 &  7222 & \textbf{\^{}12838} & 12841 & \textbf{*20389} & 20391 &  \textbf{*31132} &  31141 &  \textbf{*45677} &  45687 \\
             598a &  2398 &  2398 &  8001 &  8001 & \textbf{*15921} & 15922 & \textbf{*25694} & 25702 &  \textbf{*38576} &  38581 &  \textbf{*\^{}56094} &  56097 \\
      fe\_ocean &   464 &   464 &  1882 &  1882 &  4188 &  4188 &  7713 &  7713 &  \textbf{\^{}12667} &  12684 &  \textbf{\^{}20061} &  20069 \\
              144 &  6486 &  6486 & \textbf{\^{}15194} & 15196 & 25273 & 25273 & \textbf{*37566} & 37571 &  \textbf{*55467} &  55475 &  \textbf{*77391} &  77402 \\
             wave &  8677 &  8677 & \textbf{*17193} & 17198 & \textbf{*29188} & 29198 & \textbf{*42639} & 42646 &  \textbf{*61100} &  61108 &  \textbf{\^{}83987} &  83994 \\
             m14b &  3836 &  3836 & \textbf{*13061} & 13062 & \textbf{*25834} & 25838 & \textbf{*42161} & 42172 &  \textbf{*65469} &  65529 &  \textbf{\^{}96446} &  96452 \\
             auto & \textbf{*\^{}10101} & 10103 & \textbf{*27092} & 27094 & \textbf{*45991} & 46014 & \textbf{\^{}77391} & 77418 & \textbf{*121911} & 121944 & \textbf{\^{}172966} & 172973 \\

      \hline
    \end{tabular}

   }

    \end{table}

    \begin{table}[h!]
   \centering
    \small
      \caption{Improvement of existing partitions from the Walshaw benchmark with $\epsilon = 1\%$ using our ILP approach. In each $k$-column the results computed by our approach are on the left and the current Walshaw cuts are on the right. Results achieved by \texttt{Gain$_{\rho=-1}$} are marked with \^\ and results achieved by \texttt{Gain$_{\rho=-2}$} are marked with *. \label{table:detailedimprovement}}
\resizebox{\linewidth}{!}{

  \begin{tabular}{|l|rr|rr|rr|rr|rr|rr|}

\hline
      Graph / k & \multicolumn{2}{|c|}{2} & \multicolumn{2}{|c|}{4} & \multicolumn{2}{|c|}{8} & \multicolumn{2}{|c|}{16} & \multicolumn{2}{|c|}{32} & \multicolumn{2}{|c|}{64} \\ \hline \hline

            add20 &   585 &   585 &  1147 &  1147 &  \textbf{*\^{}1680} &  1681 &  2040 &  2040 &   2361 &   2361 &   2949 &   2949 \\
             data &   188 &   188 &   376 &   376 &   656 &   656 &  1121 &  1121 &   1799 &   1799 &   2839 &   2839 \\
             3elt &    89 &    89 &   199 &   199 &   340 &   340 &   568 &   568 &    953 &    953 &   1532 &   1532 \\
               uk &    19 &    19 &    40 &    40 &    80 &    80 &   142 &   142 &    246 &    246 &    408 &    408 \\
            add32 &    10 &    10 &    33 &    33 &    66 &    66 &   117 &   117 &    212 &    212 &    485 &    485 \\
         bcsstk33 & 10097 & 10097 & 21338 & 21338 & 34175 & 34175 & 54505 & 54505 &  77195 &  77195 & 106902 & 106902 \\
        whitaker3 &   126 &   126 &   380 &   380 &   654 &   654 &  1083 &  1083 &   1664 &   1664 &   2480 &   2480 \\
            crack &   183 &   183 &   362 &   362 &   676 &   676 &  1081 &  1081 &   1669 &   1669 &   2523 &   2523 \\
      wing\_nodal &  1695 &  1695 &  3559 &  3559 &  5401 &  5401 &  8302 &  8302 &  \textbf{*11731} &  11733 &  \textbf{*\^{}15734} &  15736 \\
        fe\_4elt2 &   130 &   130 &   349 &   349 &   603 &   603 &  1000 &  1000 &   1608 &   1608 &   \textbf{\^{}2470} &   2472 \\
         vibrobox & 10310 & 10310 & 18943 & 18943 & 24422 & 24422 & \textbf{*\^{}31710} & 31712 &  \textbf{*\^{}39396} &  39400 &  \textbf{*46529} &  46541 \\
         bcsstk29 &  2818 &  2818 &  8029 &  8029 & 13891 & 13891 & 21694 & 21694 &  34606 &  34606 &  \textbf{*\^{}54950} &  54951 \\
             4elt &   138 &   138 &   320 &   320 &   532 &   532 &   927 &   927 &   1535 &   1535 &   2546 &   2546 \\
       fe\_sphere &   386 &   386 &   766 &   766 &  1152 &  1152 &  1708 &  1708 &   2479 &   2479 &   3534 &   3534 \\
              cti &   318 &   318 &   944 &   944 &  1746 &  1746 &  2759 &  2759 &   3993 &   3993 &   5594 &   5594 \\
          memplus &  \textbf{*5452} &  5457 &  9385 &  9385 & 11672 & 11672 & 12873 & 12873 &  \textbf{\^{}13931} &  13933 &  \textbf{\^{}16091} &  16110 \\
              cs4 &   366 &   366 &   925 &   925 &  1434 &  1434 &  2061 &  2061 &   2903 &   2903 &   \textbf{\^{}3981} &   3982 \\
         bcsstk30 &  6335 &  6335 & 16583 & 16583 & 34565 & 34565 & 69912 & 69912 & 112365 & 112365 & 170059 & 170059 \\
         bcsstk31 &  2699 &  2699 &  7272 &  7272 & \textbf{*\^{}13134} & 13137 & \textbf{*23333} & 23339 &  \textbf{*37057} &  37061 &  \textbf{*57000} &  57025 \\
          fe\_pwt &   340 &   340 &   704 &   704 &  1432 &  1432 &  2797 &  2797 &   5514 &   5514 &   \textbf{\^{}8128} &   8130 \\
         bcsstk32 &  4667 &  4667 &  9180 &  9180 & \textbf{*19612} & 19624 & 35617 & 35617 & \textbf{*59501} &  59504 &  \textbf{*89893} &  89905 \\
         fe\_body &   262 &   262 &   598 &   598 &  1023 &  1023 &  1714 &  1714 &   \textbf{\^{}2748} &   2756 &   \textbf{*\^{}4664} &   4674 \\
             t60k &    75 &    75 &   208 &   208 &   454 &   454 &   805 &   805 &   1313 &   1313 &   2062 &   2062 \\
             wing &   784 &   784 &  1610 &  1610 &  2474 &  2474 &  3857 &  3857 &   \textbf{\^{}5576} &   5577 & \textbf{\^{}7585} &   7586 \\
           brack2 &   708 &   708 &  3013 &  3013 &  7029 &  7029 & 11492 & 11492 &  \textbf{*17120} &  17128 &  \textbf{\^{}25604} &  25607 \\
         finan512 &   162 &   162 &   324 &   324 &   648 &   648 &  1296 &  1296 &   2592 &   2592 &  10560 &  10560 \\
        fe\_tooth &  3814 &  3814 &  \textbf{*6843} &  6844 & 11358 & 11358 & \textbf{*\^{}17264} & 17265 &  \textbf{*24799} &  24804 &  \textbf{\^{}34159} &  34170 \\
        fe\_rotor &  2031 &  2031 &  7158 &  7158 & 12616 & 12616 & \textbf{\^{}20146} & 20152 &  \textbf{*30975} &  30982 & \textbf{*45304} &  45321 \\
             598a &  2388 &  2388 &  7948 &  7948 & 15831 & 15831 & \textbf{*25620} & 25624 &  \textbf{\^{}38410} &  38422 &  \textbf{*55867} &  55882 \\
        fe\_ocean &   \textbf{\^{}385} & 387 &  1813 &  1813 &  \textbf{*4060} &  4063 &  7616 &  7616 & \textbf{\^{}12523} &  12524 & \textbf{*19851} &  19852 \\
              144 &  \textbf{*6476} &  6478 & 15140 & 15140 &\textbf{*25225} & 25232 & \textbf{*37341} & 37347 &  \textbf{*55258} &  55277 &  \textbf{*76964} &  76980 \\
             wave &  \textbf{*\^{}8656} &  8657 & \textbf{\^{}16745} & 16747 & \textbf{*28749} & 28758 & \textbf{*42349} & 42354 &  \textbf{*60617} &  60625 & \textbf{\^{}83451} &  83466 \\
             m14b &  3826 &  3826 & 12973 & 12973 & \textbf{*\^{}25626} & 25627 & \textbf{*42067} & 42080 & \textbf{*64684} &  64697 &  \textbf{\^{}96145} &  96169 \\
             auto &  9949 &  9949 & \textbf{*26611} & 26614 & \textbf{*45424} & 45429 & \textbf{*76533} & 76539 & \textbf{*120470} & 120489 & \textbf{\^{}171866} & 171880 \\

      \hline
    \end{tabular}

   }

    \end{table}

    \begin{table}[h!]
   \centering
    \small
      \caption{Improvement of existing partitions from the Walshaw benchmark with $\epsilon = 3\%$ using our ILP approach. In each $k$-column the results computed by our approach are on the left and the current Walshaw cuts are on the right. Results achieved by \texttt{Gain$_{\rho=-1}$} are marked with \^\ and results achieved by \texttt{Gain$_{\rho=-2}$} are marked with *. \label{table:detailedimprovement}}
\resizebox{\linewidth}{!}{

  \begin{tabular}{|l|rr|rr|rr|rr|rr|rr|}

\hline
      Graph / k & \multicolumn{2}{|c|}{2} & \multicolumn{2}{|c|}{4} & \multicolumn{2}{|c|}{8} & \multicolumn{2}{|c|}{16} & \multicolumn{2}{|c|}{32} & \multicolumn{2}{|c|}{64} \\ \hline \hline

            add20 &   560 &   560 &  1134 &  1134 &  1673 &  1673 &  2030 &  2030 &   2346 &   2346 &   2920 &   2920 \\
             data &   185 &   185 &   369 &   369 &   638 &   638 &  1088 &  1088 &   1768 &   1768 &   \textbf{*2781} &   2783 \\
             3elt &    87 &    87 &   198 &   198 &   334 &   334 &   561 &   561 &    944 &    944 &   1512 &   1512 \\
               uk &    18 &    18 &    39 &    39 &    78 &    78 &   139 &   139 &    240 &    240 &    397 &    397 \\
            add32 &    10 &    10 &    33 &    33 &    66 &    66 &   117 &   117 &    212 &    212 &    476 &    476 \\
         bcsstk33 & 10064 & 10064 & 20762 & 20762 & 34065 & 34065 & 54354 & 54354 &  76749 &  76749 & \textbf{*105737} & 105742 \\
        whitaker3 &   126 &   126 &   378 &   378 &   649 &   649 &  1073 &  1073 &   1647 &   1647 &   \textbf{*2456} &   2459 \\
            crack &   182 &   182 &   360 &   360 &   671 &   671 &  1070 &  1070 &   1655 &   1655 &   \textbf{*\^{}2487} &   2489 \\
      wing\_nodal &  1678 &  1678 &  3534 &  3534 &  5360 &  5360 &  8244 &  8244 &  \textbf{*11630} &  11632 &  \textbf{*\^{}15612} &  15613 \\
        fe\_4elt2 &   130 &   130 &   341 &   341 &   595 &   595 &   990 &   990 &   1593 &   1593 &   \textbf{\^{}2431} &   2435 \\
         vibrobox & 10310 & 10310 & 18736 & 18736 & 24153 & 24153 & \textbf{*\^{}31440} & 31443 &  \textbf{*39197} &  39201 & \textbf{*46231} &  46235 \\
         bcsstk29 &  2818 &  2818 &  7971 &  7971 & 13710 & 13710 & 21258 & 21258 &  33807 &  33807 &  54382 &  54382 \\
             4elt &   137 &   137 &   319 &   319 &   522 &   522 &   901 &   901 &   1519 &   1519 &   2512 &   2512 \\
       fe\_sphere &   384 &   384 &   764 &   764 &  1152 &  1152 &  1696 &  1696 &   2459 &   2459 & \textbf{*\^{}3503} &   3505 \\
              cti &   318 &   318 &   916 &   916 &  1714 &  1714 &  2727 &  2727 &   3941 &   3941 &   \textbf{*5522} &   5524 \\
          memplus &  \textbf{*\^{}5352} &  5353 &  9309 &  9309 & \textbf{*\^{}11584} & 11586 & 12834 & 12834 &  \textbf{*13887} &  13895 &  \textbf{*15950} &  15953 \\
              cs4 &   360 &   360 &   917 &   917 &  \textbf{*\^{}1423} &  1424 &  2043 &  2043 & \textbf{*2884} &   2885 & \textbf{\^{}3979} &   3980 \\
         bcsstk30 &  6251 &  6251 & 16372 & 16372 & 34137 & 34137 & 69357 & 69357 & 110334 & 110334 & \textbf{*168271} & 168274 \\
         bcsstk31 &  2676 &  2676 &  7148 &  7148 & 12962 & 12962 & \textbf{*22949} & 22956 &  \textbf{*36567} &  36587 & \textbf{*56025} &  56038 \\
          fe\_pwt &   340 &   340 &   700 &   700 &  1410 &  1410 &  2754 &  2754 &   5403 &   5403 &   8036 &   8036 \\
         bcsstk32 &  4667 &  4667 &  8725 &  8725 & 19485 & 19485 & \textbf{*\^{}34869} & 34875 & \textbf{\^{}58739} &  58740 & \textbf{*89478} &  89479 \\
         fe\_body &   262 &   262 &   598 &   598 &  1016 &  1016 &  1693 &  1693 &   \textbf{*\^{}2708} &   2709 &   \textbf{*\^{}4522} &   4523 \\
             t60k &    71 &    71 &   203 &   203 &   449 &   449 &   792 &   792 &   1302 &   1302 &   \textbf{*\^{}2034} &   2036 \\
             wing &   773 &   773 &  1593 &  1593 &  2451 &  2451 &  \textbf{\^{}3783} &  3784 &   5559 &   5559 &   7560 &   7560 \\
           brack2 &   684 &   684 &  2834 &  2834 &  6778 &  6778 & \textbf{*11253} & 11256 &  \textbf{*\^{}16981} &  16982 &  \textbf{*\^{}25362} &  25363 \\
         finan512 &   162 &   162 &   324 &   324 &   648 &   648 &  1296 &  1296 &   2592 &   2592 &  10560 &  10560 \\
        fe\_tooth &  3788 &  3788 &  6756 &  6756 & 11241 & 11241 & \textbf{*17107} & 17108 &  \textbf{*24623} &  24625 &  \textbf{*33779} &  33795 \\
        fe\_rotor &  1959 &  1959 &  \textbf{*\^{}7049} &  7050 & 12445 & 12445 & \textbf{*19863} & 19867 &  \textbf{*30579} &  30587 &  \textbf{*44811} &  44822 \\
             598a &  2367 &  2367 &  7816 &  7816 & 15613 & 15613 & \textbf{*\^{}25379} & 25380 &  \textbf{*38093} &  38105 & \textbf{*55358} &  55364 \\
        fe\_ocean &   311 &   311 &  1693 &  1693 &  3920 &  3920 &  7405 &  7405 &  \textbf{\^{}12283} &  12288 &  19518 &  19518 \\
              144 &  \textbf{*\^{}6430} &  6432 & 15064 & 15064 & \textbf{*24901} & 24905 & \textbf{*\^{}36999} & 37003 &  \textbf{*54800} &  54806 &  \textbf{*76548} &  76557 \\
             wave &  8591 &  8591 & \textbf{\^{}16633} & 16638 & 28494 & 28494 & 42139 & 42139 &  \textbf{*60334} &  60356 &  \textbf{*82809} &  82811 \\
             m14b &  3823 &  3823 & 12948 & 12948 & 25390 & 25390 & 41778 & 41778 &  \textbf{\^{}64354} &  64364 &  \textbf{*\^{}95575} &  95587 \\
             auto &  9673 &  9673 & 25789 & 25789 & \textbf{*\^{}44724} & 44732 & \textbf{*\^{}75665} & 75679 & \textbf{\^{}119131} & 119132 & \textbf{\^{}170295} & 170314 \\

      \hline
    \end{tabular}

   }

    \end{table}

    \begin{table}[h!]
   \centering
    \small
      \caption{Improvement of existing partitions from the Walshaw benchmark with $\epsilon = 5\%$ using our ILP approach. In each $k$-column the results computed by our approach are on the left and the current Walshaw cuts are on the right. Results achieved by \texttt{Gain$_{\rho=-1}$} are marked with \^\ and results achieved by \texttt{Gain$_{\rho=-2}$} are marked with *. \label{table:detailedimprovement}}
\resizebox{\linewidth}{!}{

  \begin{tabular}{|l|rr|rr|rr|rr|rr|rr|}

\hline
      Graph / k & \multicolumn{2}{|c|}{2} & \multicolumn{2}{|c|}{4} & \multicolumn{2}{|c|}{8} & \multicolumn{2}{|c|}{16} & \multicolumn{2}{|c|}{32} & \multicolumn{2}{|c|}{64} \\ \hline \hline
            add20 &   536 &   536 &  1120 &  1120 &  1657 &  1657 &  2027 &  2027 &   2341 &   2341 &   2920 &   2920 \\
             data &   181 &   181 &   363 &   363 &   628 &   628 &  1076 &  1076 &   1743 &   1743 &   2747 &   2747 \\
             3elt &    87 &    87 &   197 &   197 &   329 &   329 &   557 &   557 &    930 &    930 &   1498 &   1498 \\
               uk &    18 &    18 &    39 &    39 &    75 &    75 &   137 &   137 &    236 &    236 &    394 &    394 \\
            add32 &    10 &    10 &    33 &    33 &    63 &    63 &   117 &   117 &    212 &    212 &    476 &    476 \\
         bcsstk33 &  9914 &  9914 & 20158 & 20158 & 33908 & 33908 & 54119 & 54119 &  \textbf{\^{}76070} &  76079 & \textbf{*105297} & 105309 \\
        whitaker3 &   126 &   126 &   376 &   376 &   644 &   644 &  1068 &  1068 &   1632 &   1632 &   \textbf{*\^{}2425} &   2429 \\
            crack &   182 &   182 &   360 &   360 &   666 &   666 &  1063 &  1063 &   1655 &   1655 &   \textbf{*\^{}2479} &   2489 \\
      wing\_nodal &  1668 &  1668 &  3520 &  3520 &  5339 &  5339 &  8160 &  8160 &  \textbf{*11533} &  11536 &  \textbf{*\^{}15514} &  15515 \\
        fe\_4elt2 &   130 &   130 &   335 &   335 &   578 &   578 &   979 &   979 &   1571 &   1571 &   \textbf{\^{}2406} &   2412 \\
         vibrobox & 10310 & 10310 & 18690 & 18690 & 23924 & 23924 & \textbf{\^{}31216} & 31218 &  \textbf{*\^{}38823} &  38826 &  \textbf{*45987} &  45994 \\
         bcsstk29 &  2818 &  2818 &  7925 &  7925 & 13540 & 13540 & 20924 & 20924 &  33450 &  33450 &  53703 &  53703 \\
             4elt &   137 &   137 &   315 &   315 &   515 &   515 &   887 &   887 &   1493 &   1493 &   \textbf{\^{}2478} &   2482 \\
       fe\_sphere &   384 &   384 &   762 &   762 &  1152 &  1152 &  1678 &  1678 &   2427 &   2427 &   3456 &   3456 \\
              cti &   318 &   318 &   889 &   889 &  1684 &  1684 &  2701 &  2701 &   3904 &   3904 &   \textbf{\^{}5460} &   5462 \\
          memplus &  \textbf{*\^{}5253} &  5263 &  \textbf{*9281} &  9292 & \textbf{*\^{}11540} & 11543 & 12799 & 12799 &  \textbf{*13857} &  13867 &  \textbf{*15875} &  15877 \\
              cs4 &   353 &   353 &   908 &   908 &  1420 &  1420 &  \textbf{\^{}2042} &  2043 &   \textbf{*2855} &   2859 &   \textbf{*\^{}3959} &   3962 \\
         bcsstk30 &  6251 &  6251 & 16165 & 16165 & 34068 & 34068 & 68323 & 68323 & 109368 & 109368 & \textbf{*166787} & 166790 \\
         bcsstk31 &  \textbf{*\^{}2660} &  2662 &  7065 &  7065 & \textbf{*\^{}12823} & 12825 & \textbf{*22718} & 22724 &  \textbf{*36354} &  36358 &  \textbf{*55250} &  55258 \\
          fe\_pwt &   340 &   340 &   700 &   700 &  1405 &  1405 &  2737 &  2737 &   \textbf{\^{}5305} &   5306 &   \textbf{\^{}7956} &   7959 \\
         bcsstk32 &  4622 &  4622 &  8441 &  8441 & 18955 & 18955 & 34374 & 34374 &  58352 &  58352 &  \textbf{*88595} &  88598 \\
         fe\_body &   262 &   262 &   588 &   588 &  1012 &  1012 &  1683 &  1683 &   \textbf{*\^{}2677} &   2678 &   \textbf{\^{}4500} &   4501 \\
             t60k &    65 &    65 &   195 &   195 &   441 &   441 &   787 &   787 &   \textbf{*1289} &   1291 &   \textbf{*\^{}2013} &   2015 \\
             wing &   770 &   770 &  \textbf{*1589} &  1590 &  2440 &  2440 &  3775 &  3775 &   \textbf{*\^{}5512} &   5513 &   \textbf{\^{}7529} &   7534 \\
           brack2 &   660 &   660 &  2731 &  2731 &  6592 &  6592 & \textbf{*11052} & 11055 &  16765 &  16765 &  \textbf{*25100} &  25108 \\
         finan512 &   162 &   162 &   324 &   324 &   648 &   648 &  1296 &  1296 &   2592 &   2592 &  10560 &  10560 \\
        fe\_tooth &  3773 &  3773 &  6687 &  6687 & \textbf{*\^{}11147} & 11151 & \textbf{*16983} & 16985 &  \textbf{\^{}24270} &  24274 &  \textbf{*33387} &  33403 \\
        fe\_rotor &  1940 &  1940 &  6779 &  6779 & \textbf{*12308} & 12309 & \textbf{*19677} & 19680 &  \textbf{*30355} &  30356 &  \textbf{*44368} &  44381 \\
             598a &  2336 &  2336 &  \textbf{*7722} &  7724 & 15413 & 15413 & 25198 & 25198 &  \textbf{\^{}37632} &  37644 &  \textbf{*54677} &  54684 \\
        fe\_ocean &   311 &   311 &  1686 &  1686 &  3886 &  3886 &  7338 &  7338 &  \textbf{\^{}12033} &  12034 &  \textbf{*\^{}19391} &  19394 \\
              144 &  6345 &  6345 & \textbf{\^{}14978} & 14981 & \textbf{*24174} & 24179 & \textbf{*\^{}36608} & 36608 &  \textbf{*54160} &  54168 &  \textbf{*75753} &  75777 \\
             wave &  8524 &  8524 & \textbf{*16528} & 16531 & 28489 & 28489 & \textbf{*\^{}42024} & 42025 &  \textbf{*\^{}59608} &  59611 &  \textbf{*81989} &  82006 \\
             m14b &  3802 &  3802 & \textbf{*\^{}12858} & 12859 & 25126 & 25126 & \textbf{*41097} & 41098 &  \textbf{*63397} &  63411 &  \textbf{*94123} &  94140 \\
             auto &  9450 &  9450 & 25271 & 25271 & 44206 & 44206 & \textbf{*74266} & 74272 & \textbf{*118998} & 119004 & \textbf{\^{}169260} & 169290 \\

      \hline
    \end{tabular}

   }

    \end{table}

\end{appendix}

\end{document}